\documentclass{ws-jai}
\usepackage[flushleft]{threeparttable}
\usepackage{comment}
\usepackage{natbib}
\newcommand\farcm{\mbox{$.\mkern-4mu^\prime$}}%
\newcommand\arcsec{\mbox{$^{\prime\prime}$}}%
\newcommand\arcmin{\mbox{$^{\prime}$}}%

\newcommand\micron{\mbox{$\mu$m}}
\newcommand\apj{ApJ}%
\newcommand\apjs{ApJS}%
\newcommand\aap{A\&A}%
\newcommand\procspie{Proc.~SPIE}%

\begin{document}

\catchline{}{}{}{}{} 

\markboth{Madhwani et al.}{TIFR CSU for multi-object spectroscopy}

\title{A compact cryogenic configurable slit unit for a multi-object infrared spectrograph:\\
Design and Development of a prototype at TIFR
}

\author{P. P. Madhwani, A. P. K. Kutty, B. Mookerjea$^a$, J. V. Parmar, V. N. Kurhade, S. L. D'Costa, P. Manoj, A. Surya}

\address{
Department of Astronomy \& Astrophysics, Tata Institute of Fundamental Research, \\ Homi Bhabha Road, Mumbai 400005, India\\
}

\maketitle

\corres{$^a$Corresponding author: B. Mookerjea\, bhaswati@tifr.res.in}

\begin{history}
\received{(to be inserted by publisher)};
\revised{(to be inserted by publisher)};
\accepted{(to be inserted by publisher)};
\end{history}

\begin{abstract}
We present a cryogenic configurable slit unit (CSU) for a multi object infrared spectrograph with an effective field of view of 9\farcm1$\times$9\farcm1 that was completely conceived and designed in the laboratory at TIFR. Several components of the CSU including the controller for the commercially procured piezo-walkers, controlled loop position sensing mechanism using digital slide callipers and a cryogenic test facility for the assembled prototype were also developed in-house. The principle of the CSU involves division of the field of view of the spectrometer into contiguous and parallel spatial bands, each one associated with two opposite sliding metal bars that can be positioned to create a slit needed to make spectroscopic observations of one astronomical object. A three-slit prototype of the newly designed CSU was built and tested extensively at ambient and cryogenic temperatures. The performance of the CSU was found to be as per specifications.
\end{abstract}


\section{Introduction}

\noindent
Near-infrared (NIR) spectroscopy is particularly well-suited for observation of vibrational and rotational molecular lines that are indicative of the chemo-dynamical structure of stars of the Milky Way and can be used to study stellar radial velocities and abundances  \citep{Wallace1997}. The NIR spectra of young stellar objects and stars provide important constraints on the accretion of material, the structure of protostellar disks that eventually form planets \citep{Scoville1983, Alcala2014}. Spectroscopic observations in the NIR can be used to access the high redshift galaxies whose important rest frame optical diagnostic emission lines are redshifted to the NIR \citep{Petitjean2011}. 

Multi Object Spectroscopy (MOS)is an efficient way to observe the spectra of an aggregate of sources that are within the field of view of an instrument at a single pointing. With the advent of technologies enabling fabrication of large format detector arrays, the demand for capabilities to perform simultaneous spectroscopy of multiple objects within the fields of view (FOV) covering  several square arcminutes up to several square degrees of the large telescopes has been on the rise  \citep{Ellis1988,Allington2006}. 

In the instruments capable of MOS, the light from the source is focused by the telescope onto the focal plane, where narrow slits or fiber optics are used to isolate the source light and is finally dispersed using a spectrograph with a dispersing element like a prism or a diffraction grating.  The spectra are then imaged onto a detector. The data undergo detector level processing  and extraction, as well as wavelength and flux calibration, typically using software designed specifically to handle the multiplexed spectra. Presently, most large 8–10 meter class ground-based observatories offer at least one MOS instrument option. In ground-based instruments, apertures are either created using slit masks or using fibers arranged into patterns on the sources in the field. For optical telescopes the field-specific slit masks are milled out on-site prior to the observations. Such pre-fabricated slit masks are inconvenient in the near-infrared (1--2\,\micron) where there is a strong contribution from background thermal emission that necessitate cooling the slits to temperatures of 120\,K or below along with the entire spectrograph. A plausible solution for such instruments is the use of a configurable slit unit (CSU) to precisely position the slit mask that can be configured in-situ in the cryogenically cooled environment shortly before observations begin. Configurable slit spectrographs have a number of benefits over slit mask-based spectrographs, that include flexibility, efficiency (easy to customize the slit width based on the properties of the target) and reduced overhead (in-situ configuration in cryogenic environment). Such configurable slit units are currently installed on the MOSFIRE instrument on the Keck telescope \citep{mclean2010} and also on Espectr\'ografo Multiobjeto Infra-Rojo (EMIR) instrument on the Gran Telescopio de Canarias (GTC). MOSFIRE can observe up to 46 objects in a 6\farcm 12$\times$6\farcm 12 FOV and EMIR on GTC can observe up to 53 objects in a FOV of 6\farcm 64$\times$4\arcmin\ \citep{garzon2022}. The physical size of the EMIR FOV  in the focal plane is 307\,mm$\times 307$\,mm and for the MOSFIRE the FOV is 267\,mm$\times 267$\,mm.

We present here the design of a configurable slit unit operational at cryogenic temperatures (120\,K) meant for a Multi Object Infrared Spectrograph (MOIS) \citep{Surya2023} operational between 0.8--2.5\,\micron, to be installed on an Indian optical/near-infrared observing facility. The present prototype for the CSU has 5 configurable slits designed as per the optics of the 3.6\,m Devasthal Optical Telescope (DOT). However the modular and scalable nature of the design will enable installation of the CSU on future 10\,m class Indian telescopes that are being planned. The entire design has been developed in-house and a prototype consisting of three slits has already been fabricated and tested at cryogenic temperatures.

\section{Design of the CSU}

The design goals for the CSU was to create 5 slits using 10 bars each of which can have a maximum displacement of up to 100\,mm for a field of view of  85.77\,mm$\times 85.77$\,mm. The position of the slit in the FOV and its width are decided upon based on the location and size of the astronomical source. The CSU is required to generate a multi-slit configuration, a long slit, or an imaging aperture while being at cryogenic temperatures, necessary for its operation at the infrared wavelengths.


\begin{figure}[h]
\centering
\includegraphics[width=0.7\textwidth]{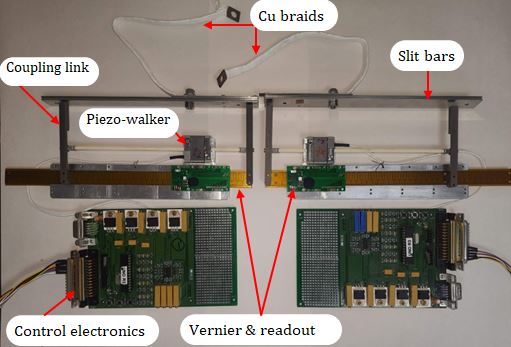}
\caption{Single slit pair assembly along with the drive electronics. The functional components of the bar assembly are labeled.
\label{fig_slitbar}}
\end{figure}
\noindent 
Each slit bar assembly for the CSU consists of i) Vernier scale, ii) Piezo-walker,iii) Piezo-walker mounting base, iv) Titanium link rods, v) Slit bar and Slit edge and the vi) Controller electronics (Figure\,\ref{fig_slitbar}). In the following text we present details of the crucial components of this assembly.

\subsection{Piezo-walker based drive mechanism}

At the operating wavelengths of the MOIS it is necessary to suppress the background emission from the components of the spectrometer, hence the slit bars need to be in vacuum and in a cryogenic environment. This requirement of low temperatures impose major constraints on the drive mechanisms that can be used for the motion of the slit bars. Additionally, it is preferable that the power dissipation from the bars and the driving motors is negligible when the bars are not in motion. Based on these considerations we have identified piezo-walkers as the drivers for the slit bars, which further have the advantage of negligible backlash error as compared to regular rotary stepper motors that are also used for such purposes. The PI NEXACT-310 piezo-walkers identified for this purpose provide a travel range 10 to 125\,mm and are capable of generating a force of up to 10\,N. The piezo-walker drives a ceramic runner which was mechanically interfaced with the slit bar through a titanium link rod and the motion is controlled by a controller designed in-house. The NEXACT-310 piezo-walkers are functional only down to a temperature of 273\,K, hence the piezo-walkers were isolated from thermal radiation using enclosures made out of space-grade multi-layer insulating (MLI) sheet. Only the slit bars were cooled using copper braids connected to a cold plate at liquid N$_2$ temperature of 77\,K (Fig.\ref{fig_slitbar}).

\subsection{Closed-loop Vernier-scale based accurate Positioning of the Slit Bars}

Accurate positioning of the slit bars was achieved using Grid-Capacitance Linear Synchronous Sensors which consist of a set of grid-capacitances and signal processing circuit similar to those used in commercially available digital vernier calipers. The position information was tapped from the readout to enable closed-loop correction to achieve an absolute positioning accuracy of 10\,\micron\ for each slit bar. The positioning mechanism was also designed and developed in-house. The slit length is fixed (17.15\,mm) and the width can range from 0 (fully shut) to 90\,mm (fully open). The slit positioning accuracy is $\pm 10$\,\micron\ and the slit width accuracy is $\pm 14$\,\micron. The design goal is a slit width of 120\,\micron. The motion of the slit-bars is restricted within the desired range by using limit switches, which when depressed stop the movement in a particular direction and subsequently only motion in the reverse direction is possible. The positioning accuracy of the CSU was tested by shinning light through the assembly and the slits of different widths that were formed at different positions of the FOV were imaged using a high resolution camera fitted to a microscope. The attained positions and widths of the slits were accurate to within 15\,\micron\ and were found to be consistently reproducible.

\subsection{Mechanical Design of the Slit Bars}
\begin{figure}[h]
\centering
\includegraphics[width=0.3\textwidth]{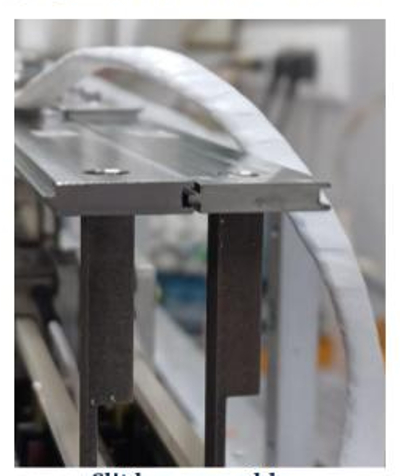}
\caption{Cross-section of the single slit bar designed to allow smooth motion without light leak. The copper braids used to cool the slit bars are also shown.
\label{fig_barcross}}
\end{figure}

The mechanical design of the slit bars was challenging since the movement of the slit bars had to be smooth while ensuring that there is no light leak through them.  In order to meet both design goals, a special design for the cross-section of the slit bars was devised (Fig.\,\ref{fig_barcross}) which ensured complete overlap of the bars at their edges as well as free movement.  Each slit bar consists of two parts the main body or the bar and the edge. The bar, made of stainless steel has  poor thermal conductivity. The edge of the slit bar is made from a hard aluminium alloy plate with a thickness of 800\,\micron\ attached to an overlapping machined edge having a thickness of 300\,\micron. The choice of aluminium alloy for the slit edge is to ensure efficient cooling to cryogenic temperatures using copper braids connected to the cold plate of the liquid N$_2$ cylinder. The titanium link rods which were machined as per the precision requirement play an important role in the assembly because these isolate the slit bars from the rest of the assembly thus minimizing the thermal load on the copper braids. The titanium rod in the slit bar assembly works as a link between the Vernier scale, the piezo-walker and the slit bar.

\subsubsection{Controller Electronics}

\begin{figure}[h]
\centering
\includegraphics[width=0.8\textwidth]{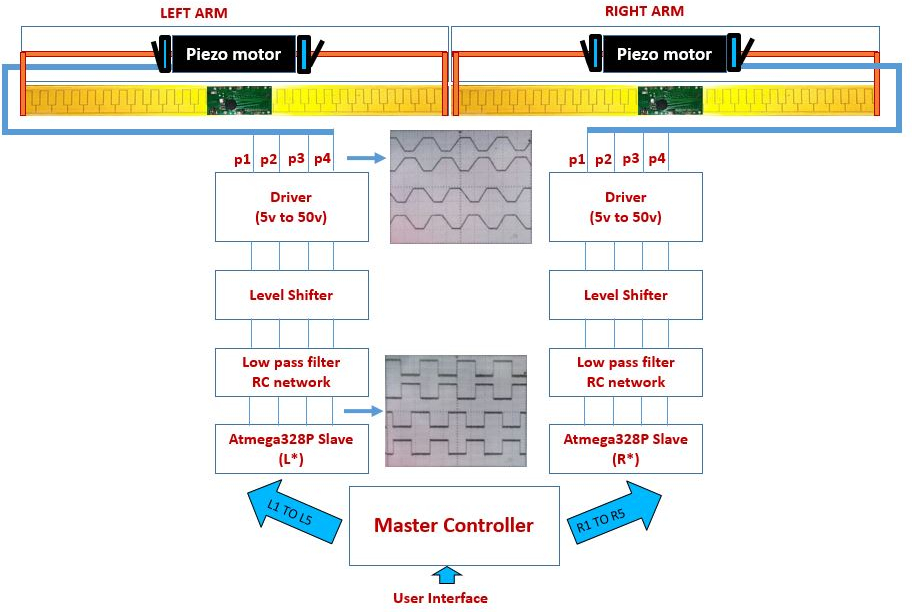}
\caption{Schematic to show the working principle of the functional blocks of the controller electronics for the piezowalkers. 
\label{fig_drive}}
\end{figure}

The proto type CSU model developed  comprises of  3 pairs of  slit bar assemblies and each slit bar assembly has a  left arm and a right arm. The  runner of the piezo walker is coupled to the slit bar as well as a  electronic Vernier scale which provides the position information with an accuracy of 10\,\micron. Each of the six piezo walkers is driven individually by a  drive controller consisting of closed loop circuits developed in-house. The working principle of the control circuit can be understood from the functional block diagram of 1 slit assembly (Fig.\,\ref{fig_drive}). In the following although the description is for the three slit assembly for which the mechanical parts have been fabricated, the electronics has already been designed and built and tested for the five-slit final CSU configuration.

The piezo walker has four legs which  when excited with a 50\,V trapezoidal wave in a particular sequence produces a linear motion in the runner. The controller circuit is built around a microcontroller which generates the square waves of desired frequency, which are then converted to sine waves by passing through  low pass filters. The signals are further amplified, processed and level shifted to achieve 4 trapezoidal waveforms of 50\,V as required to drive the four legs of the piezo walker. The functional block represents a single slit assembly. Two more such assemblies together complete the three slit assembly. These three slit assemblies termed as slaves are controlled through a master which is commanded through a user interface.

Operationally, when powered up all the 6 slit bars are moved to the retracted position so that the FOV of the detector is exposed and when fully open, the limit switch is depressed to set the Vernier readings to a reference value. At the engineering model level, using the user interface it is possible to provide the location and size of all 3 slits, by providing  6 parameters i.e the 3 positions corresponding to the location of the left arm of each slit and 3 slit widths. The parameters thus supplied are used by the master controller to compute the value of distances to be traversed by the 6 arms of the CSU. The computed values are then sent to the corresponding slaves, which enable the movement of the slit bars by the piezos till the desired position and width of all slits are achieved. The configuration of slits on the FOV so achieved remain unchanged till a new set of parameters is transmitted by the user to the master controller. The master controller stores the current location of the slits, so that when a new set of parameters is entered it calculates the range of motion required for each slit bar relative to that position. The controller is programmed to enable motion of the slit bars with a coarse and a fine step depending on the distance between the actual and the desired positions of the bars. The controller is also designed and programmed to avoid accidental collisions of the slit bars.

\begin{figure}[h]
\centering
\includegraphics[width=0.9\textwidth, keepaspectratio]{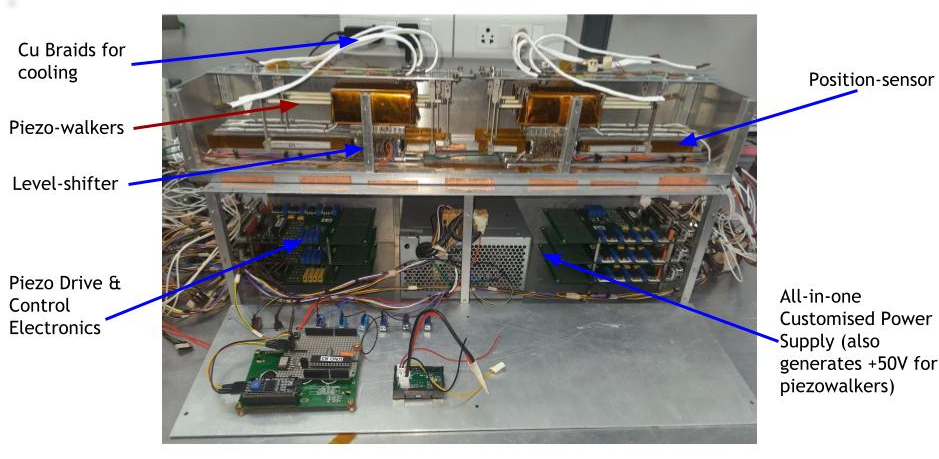}
\caption{Completed CSU prototype creating 3-slits at the FOV in the laboratory. The slit bar assembly on the top will be inside the cryogenic dewar while the drive and monitoring electronics will be outside. 
\label{fig_csu}}
\end{figure}

Figure\,\ref{fig_csu} shows the 3-slit CSU prototype being tested in the laboratory. In the final configuration the drive electronics consisting of ten cards will be kept outside of the cryostat and will be used to control the movement of the slits, monitor the positions and temperatures of the different parts of the CSU through hermetically sealed connectors.

\section{Cryogenic Functionality Testing of the CSU in the laboratory}
\begin{figure}[h]
\centering
\includegraphics[width=0.35\textwidth, keepaspectratio]{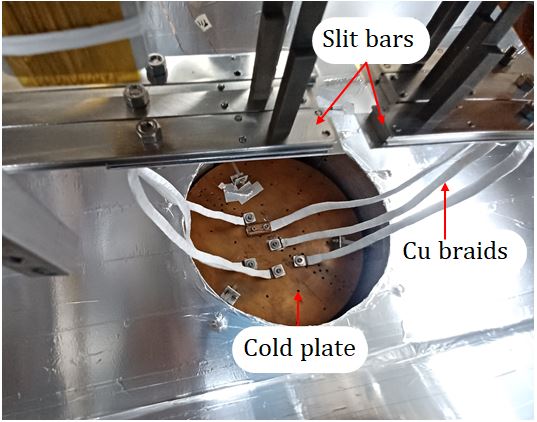}
\includegraphics[width=0.6\textwidth, keepaspectratio]{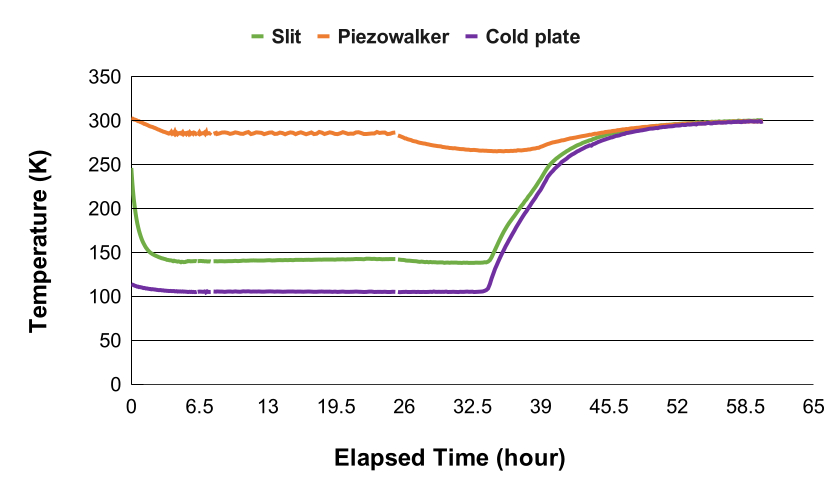}
\caption{({\em Left}) The CSU connected to the cold plate (at the bottom of the liquid N$_2$ tank) inside the dewar prior to the cryogenic testing of the unit in the laboratory. The shinning surface around the cold plate is a radiation shield created to ensure efficient cooling by reflecting the heat due to the uncooled parts of the dewar. ({\em Right}) Typical temperature profiles of the slit bar, the body of the piezo-walker and the coldplate measured during one of the cryogenic tests in the laboratory.
\label{fig_dewar}}
\end{figure}

A dedicated aluminium dewar (supporting a vacuum of up to 10$^{-6}$\,mbar) was designed and fabricated at TIFR for the cryogenic testing of the CSU. The dewar consists of a top plate with a flange with four hermetically sealed connectors each with 19 pins as well as an inlet for the 3 litre liquid N$_2$ tank inside. The inlet pipe to the liquid nitrogen tank was fitted with a solenoid valve based system for automated round-the-clock filling (without spillage) of N$_2$ that was developed in-house. The base of the slit bar assembly was mounted on the bottom plate of the dewar with teflon rings to isolate the two plates thermally. For ease of mounting the dewar was suspended upside-down and the base plate with the CSU attached to it was attached to the dewar (Fig.\,\ref{fig_dewar}). Multiple tests of the performance of the cooling scheme as well as copper braids of different sizes (and hence thermal efficiencies) were carried out in the laboratory using this set-up. No major issues with the motion of the CSU in the final design presented here was noticed during the cryogenic tests. We present the results of one of the  tests which  was continued uninterrupted for over 72 hours and the motion of the slit bars measured at regular intervals was found to be as per specifications for the entire duration of the test. Figure\,\ref{fig_dewar} shows the results of the temperature profiles of the slit, the piezo-walker and the coldplate measured using thermistors affixed at different parts of the CSU. In order to maintain the piezo-walkers at a temperature above 273\,K it was locally heated in an automated manner. The temperatures measured in the rest of the CSU do not reflect any changes due to this heating, which suggests that the MLI enclosure of the piezo-walkers may be adequate. A more detailed analysis of the thermal background is currently in progress. 

\section{Summary and Future Directions}

We presented here the new design of a configurable slit unit for a multi-object infrared spectromgraph. The current technical specifications are based on the optics of the DOT and for a five-slit unit. However, the modular nature of the CSU design makes it easily scalable for much larger FOV spectrometers meant for bigger (10\,m or larger) telescopes and larger number of slits. The mechanical and electronic design of the CSU presented here  were developed completely indigenously. Except for the commercially procured piezo-walkers all mechanical and electronic  parts were fabricated/built in-house for the  three-slit prototype of the CSU presented here. Based on a series of tests performed in the laboratory we conclude that CSU performs as per specification both at room and cryogenic temperatures. Keeping in mind the constraint imposed by the piezo-walkers being non-functional at  temperatures below 273\,K, we are working on a design using cryogenic piezo-drives. A definite conclusion on whether a cryogenic piezo-drive is really needed will become clearer only after the opto-mechanical design of the spectrometer including the thermal modelling is completed. In this context we point out that the commercially available cryo piezodrives typically have travel ranges of only up to 50\,mm, whereas for our design a movement range up to 100\,mm is needed. Thus specially customized commercially available piezo-drives are currently being explored.

\section*{Acknowledgments}
The project was entirely funded by Department of Atomic Energy, Government of India, under Project Identification No. RTI 4002. The authors acknowledge the contributions of G. S. Meshram towards the initial design of the slit bar edges and also the support and help received from H. Shah, R. Jadhav, S. Poojary, S. B. Bhagat and S. Gharat of the IR Astronomy group, TIFR.
 


\begin{thebibliography}{9}
\bibitem[Alcal{\'a} et al.(2014)]{Alcala2014} Alcal{\'a}, J.~M., Natta, A., Manara, C.~F., et al.\ 2014, \aap, 561, A2. doi:10.1051/0004-6361/201322254
\bibitem[Allington-Smith(2006)]{Allington2006} Allington-Smith, J.\ 2006, New Astronomy Reviews, 50, 244. doi:10.1016/j.newar.2006.02.024
\bibitem[Ellis \& Parry(1988)]{Ellis1988} Ellis, R.~S. \& Parry, I.~R.\ 1988, Instrumentation for Ground-Based Optical Astronomy, 192
\bibitem[Garz{\'o}n et al.(2022)]{garzon2022}Garz{\'o}n, F., Balcells, M., Gallego, J., et al.\ 2022, \aap, 667, A107. doi:10.1051/0004-6361/202244729
\bibitem[McLean et al.(2010)]{mclean2010} McLean, I.~S., Steidel, C.~C., Epps, H., et al.\ 2010, Proceedings of the SPIE, 7735, 77351E. doi:10.1117/12.856715
\bibitem[Petitjean(2011)]{Petitjean2011} Petitjean, P.\ 2011, Astronomische Nachrichten, 332, 309. doi:10.1002/asna.201111544
\bibitem[Scoville et al.(1983)]{Scoville1983} Scoville, N., Kleinmann, S.~G., Hall, D.~N.~B., et al.\ 1983, \apj, 275, 201. doi:10.1086/161526
\bibitem[Surya et al.(2023)]{Surya2023} Surya, A., Manoj, P., Mookerjea, B., accepted for \procspie\ "Astronomical Optics: Design, Manufacture, and Test of Space and Ground Systems IV", August 2023.
\bibitem[Wallace \& Hinkle(1997)]{Wallace1997} Wallace, L. \& Hinkle, K.\ 1997, \apjs, 111, 445. doi:10.1086/313020



\end{thebibliography}
\end{document}